\documentstyle[sprocl]{article}
\def\Journal#1#2#3#4{{#1} {\bf #2}, #3 (#4)}

\def\NPB{{\em Nucl. Phys.} B}
\def\PLB{{\em Phys. Lett.}B}

\def\be{\begin{equation}}
\def\ee{\end{equation}}
\def\bea{\begin{eqnarray}}
\def\eea{\end{eqnarray}}

\begin{document}

\begin{flushleft}KCL-TH-98-57\hskip69mm hep-th/9811177\end{flushleft}\vskip4mm

\title{Brane Dynamics and Four-Dimensional Quantum Field Theory
\footnote{Talk
given by P.C. West at the Trieste Conference on
Superfivebranes and Physics in 5+1 Dimensions, April 1998.}
}
\author{N.D. Lambert and P.C. West}
\address{Department of Mathematics\\
King's College    
The Strand, London\\
WC2R 2LS, UK}
%
%

\maketitle
\abstracts{We review the relation between the classical dynamics of
the M-fivebrane and the quantum low energy effective action for 
$N=2$ Yang-Mills theories. We also discuss some outstanding issues in this
correspondence.}

\section{Introduction}

In recent years there have been some remarkable and surprising advances in
non-perturbative gauge theory arising from the study of branes in string
theory and M-theory. While there have been
many interesting developments, here we will only 
review the precise connection
between the classical dynamics of the M-fivebrane and  
four-dimensional $N=2$ quantum Yang-Mills theories. Let us begin by reviewing
some of Witten's analysis of 
type IIA brane configurations~\cite{W}. A similar role for the M-fivebrane 
also appeared in~\cite{KLMVW}.

We start by considering type IIA string theory. We place two parallel
NS-fivebranes in the $x^0,x^1,x^3,...,x^5$ plane separated along the $x^6$
direction by a distance $\Delta x^6$. These two NS-fivebranes will preserve
sixteen of the thirty-two spacetime supersymmetries. Next we introduce
$N_c$ parallel D-fourbranes in the $x^0,x^1,x^2,x^3,x^6$ plane. These
D-fourbranes stretch between the two NS-fivebranes and reduce the number
of preserved supersymmetries to eight. 

At weak string coupling the NS-fivebranes are heavy and their motion can
be ignored. The low energy fluctuations of this system are then described
by the D-fourbranes. As is well known the low energy dynamics of $N_c$
parallel D-fourbranes is given by a five-dimensional $U(N_c)$ gauge
theory with sixteen supersymmetries. 
The presence of the NS-fivebranes has two effects. 
Firstly they reduce the number of preserved supersymmetries 
to eight. Secondly,  since the $x^6$ direction of the D-fourbrane is finite
in extent, at low energy their worldvolume is four-dimensional. An
overall $U(1)$ factor of the gauge group $U(N_c)$ is trivial and simply
describes the centre of mass motion of the D-fourbranes so we may ignore it. 
Thus the
low energy, weak coupling description of this configuration is given by
four-dimensional $N=2$  $SU(N_c)$ gauge theory. 

One can also consider adding $N_f$ 
semi-infinite D-fourbranes to this configuration. These intersect
the left or right NS-fivebrane at one end but extend to infinity at the
other. Since they are infinitely heavy as compared to the finite D-fourbranes
their motion is suppressed. However in the D-brane picture there are 
stretched open strings with one end on a semi-infinite D-fourbrane and the
other on a finite D-fourbrane. These strings give rise to $N_f$ massive 
hyper-multiplets in the (anti-) fundamental representation of
$SU(N_c)$ in the four-dimensional gauge theory. Their bare mass 
given by the length of the open strings,  which is the distance between the
finite and semi-infinite D-fourbranes.

What Witten noticed was that there is an elegant strong coupling description
of this configuration in M-theory. Increasing the string coupling
lifts us  up to eleven dimensions and introduces another coordinate $x^{10}$
which is periodic with period $2\pi R$. 
Furthermore one can go to strong coupling keeping the curvatures small, 
so that supergravity is a good approximation, yet also keeping the Yang-Mills
coupling constant fixed~\cite{W}.
The NS-fivebranes simply lift to
M-fivebranes. The D-fourbranes also lift to M-fivebranes, only wrapped on
the $x^{10}$ dimension. Thus in eleven dimensions the entire configuration
appears as intersecting M-fivebranes. An important realisation is that
this configuration can be viewed as a single
M-fivebrane wrapped on a two-dimensional manifold, embedded the
four-dimensional space with coordinates $x^4,x^5,x^6,x^{10}$.

The condition that an M-fivebrane wrapped around a manifold breaks
only of half the supersymmetry, leaving
eight unbroken supersymmetries, is that the manifold is
complex~\cite{BBS}, i.e. it
must be a Riemann surface. 
It is helpful then to introduce the complex notation
\be
s = (x^6 + ix^{10})/R\ ,\quad t = e^{-s}\ ,\quad z = x^4+ix^5\ .
\ee
Thus the supersymmetric intersecting M-fivebrane configuration can be
realised by any holomorphic embedding $F(t,z)=0$ of its worldvolume into
the $x^4,x^5,x^6,x^{10}$ dimensions of eleven-dimensional spacetime.

Let us return to the configuration in question. 
At a given point $z$ there should be two solutions for $s$, corresponding
to the two NS-fivebranes. Therefore we take  $F$ to be second order in $t$
\be
A(z)t^2 -2B(z)t + C(z) =0 \ .
\ee
We expect that as $z\rightarrow\infty$, we are on either the left or right
NS-fivebrane so that $t\rightarrow\infty,0$ respectively. If $A(z)$ or $C(z)$
has a zero at any finite value of $z$, so that $t \rightarrow\infty,0$ there, 
this can be interpreted as a semi-infinite D-fourbrane ending on 
the left or right NS-fivebrane respectively. Let us just consider
semi-infinite D-fourbranes ending on the right NS-fivebrane. In this case
we may set  $A=1$ by rescaling $t$. For $N_f$ semi-infinite 
D-fourbranes we need $N_f$ zeros of $C(z)$. Thus $C(z)$ must take the form
\be
C(z) = \Lambda\prod_{a=1}^{N_f}(z - m_a)\ , 
\ee
where $\Lambda$ is a constant and $m_a$ are the positions of the semi-infinite
D-fourbranes, i.e. their bare masses. For $N_f=0$ we simply take $C=\Lambda$.
Finally we need to determine $B(z)$. For a fixed $s$ we need there to be
$N_c$ solutions for $z$, corresponding to the $N_c$ finite D-fourbranes. 
Thus $B(z)$ must take the form
\be
B(z) = \prod_{i=1}^{N_c}(z - e_i)\ ,
\ee
note that we can set the coefficient of $z^{N_c}$ to one by rescaling $z$.
For large $z$ the $e_i$ then appear as the positions of the $N_c$
finite D-fourbranes. Since we have frozen out the centre of mass motion
we set $\sum_{i=1}^{N_c}e_i=0$, which can also be achieved by redefining $z$. 
With these
conditions imposed ones see that $s(z)$ defines  precisely
the  Seiberg-Witten curve
\be
y^2 = \left(\prod_{i=1}^{N_c}(z - e_i)\right)^2 - 
\Lambda\prod_{a=1}^{N_f}(z - m_a)\ ,
\label{curve}
\ee
where $y = t-B$.

In summary then this brane configuration has a weak coupling description as 
four-dimensional $N=2$ $SU(N_c)$ Yang-Mills theory with $N_f$ hyper-multiplets
in the fundamental representation and a strong coupling description as
an M-fivebrane wrapped around the correct Seiberg-Witten curve. 
Thus from the brane configuration we can identify
the Riemann surfaces that are known to be  
associated with the exact quantum low 
energy effective action of four-dimensional $N=2$ Yang-Mills theory \cite{SW}. 
This analysis
also suggests why the scalar modes in the Seiberg-Witten solution 
correspond to moduli
of a Riemann surface, since the zero modes of the M-fivebrane are
just the Riemann surface moduli. 
In addition Witten was able to derive the appropriate curves for many
new classes of Yang-Mills theories which were previously unknown.

This remarkable correspondence left open the question as to 
{\it whether or not the 
classical M-fivebrane could predict the precise 
perturbative and instanton corrections of the Yang-Mills Theory and not just 
the Seiberg-Witten curve}. In other words, a knowledge of the elliptic
curve alone is not enough to compare with the four-dimensional Yang-Mills
quantum field theory. One must also know how to calculate the low energy
effective action from the M-fivebrane dynamics, including all the instanton
corrections.

\section{Brane Dead}

Since the paper~\cite{W} first appeared there 
have been several discussions of how to 
construct the low energy effective action 
from M-theory 
but many papers present  a  seriously flawed argument, apparently
based on misinterpretations of comments in~\cite{W}. It 
would be invidious to reference all of these articles here, however the 
following argument, which has appeared in a substantial review article, 
illustrates many of these issues. 

The  M-fivebrane worldvolume theory has   
a self-dual three-form $H$. The argument states that the 
action therefore contains the standard kinetic 
term for $p$-form fields
\begin{equation}
S_{SYM}=\int d^6 x H^2\ .
\label{action}
\end{equation}
To obtain the effective action one must decompose $H$ in a basis of
non-trivial one-forms $\Lambda_i$ of the Riemann surface $\Sigma$ (the 
Seiberg-Witten curve), 
with genus $N_c-1$, $I=1,...,N_c-1$,
\begin{equation}
H=\sum_{I=1}^{N_c-1}F_I\wedge \Lambda_I+
*F_I\wedge *\Lambda_I\ .
\label{Hdecomp}
\end{equation} 
Substituting eq. \ref{Hdecomp} 
into eq. \ref{action} leads to
\begin{equation}
S_{SYM}=\int d^4x ({\rm Im}\tau_{IJ})F_I\wedge*F_J+
({\rm Re}\tau_{IJ}) F_I\wedge F_J\ ,
\end{equation}
with $\tau$ the period matrix of $\Sigma$
\bea
{\rm Im}\tau_{IJ}&=&\int_\Sigma\Lambda_I\wedge
*\Lambda_J+c.c. \ ,\label{Imtau} \\
{\rm Re}\tau_{IJ}&=&\int_\Sigma\Lambda_I\wedge
\Lambda_J+c.c.\ .\label{Retau} \\
\nonumber
\eea

It should not take the alert reader much time to realise the following
errors in the above argument. Firstly, and most seriously, since 
the ansatz in eq. \ref{Hdecomp}  ensures $H$ is self-dual
the action eq. \ref{action} {\it vanishes identically}. This can be easily seen
since $H^2 = H\wedge *H = H\wedge H$. Now $H$ is an odd form so $H\wedge H=0$.
Therefore this argument predicts that the Seiberg-Witten effective action 
vanishes identically. 

To avoid this problem some articles reference the 
paper~\cite{V} where a first order 
six-dimensional action is given for a self-dual three form which is non-zero. 
However, this action is not coupled to scalars so that
the resulting four-dimensional action contains a
constant period matrix $\tau_{ij}$. Therefore this method leads only
to the classical low energy effective action with no quantum corrections. 
Even if one ignored this problem and let the Riemann surface
moduli $e_i$ become spacetime dependent, one could not arrive at the 
correct low energy
effective action because the relations between the moduli 
$e_i$ and the Yang-Mills scalar fields $a_I$ are still unknown. 

Given that this argument starts with zero one might wonder how its 
advocates obtain a non-trivial answer. In fact the expressions
eq. \ref{Imtau} and eq. \ref{Retau} are incorrect. To see this one only
needs to consider the case of genus one ($N_c$=2) and let us choose our 
basis of
one forms so that $\Lambda_1 = \Lambda$ is a  holomorphic 
one form,  $*\Lambda = i\Lambda$ and $\Lambda_2$ is an anti-holomorphic 
one-form, ${\bar \Lambda}$, $*{\bar \Lambda} = -i{\bar \Lambda}$. 
Then one finds that the two equations \ref{Imtau} and \ref{Retau} are the same
(up to a constant). Furthermore no redefinition will alter this since, 
in complex notation, it is clear
that the only independent integrand one could write down is 
$\Lambda\wedge{\bar \Lambda}$ and this is 
purely imaginary, i.e. only
${\rm Im}\tau$ has a simple integral formula. 

\section{The Fivebrane Equations of Motion}

Let us now discuss in detail the worldvolume theory of the M-fivebrane. It 
has a six-dimensional $(2,0)$
tensor multiplet of
massless fields on its worldvolume. The component fields of this
supermultiplet are five real scalars $X^{a'}$, a gauge field 
$B_{m n}$
whose field strength satisfies a modified self-duality condition and
sixteen spinors $\Theta ^i_\beta$. The scalars are the
coordinates  transverse to the fivebrane and correspond to the
breaking of eleven-dimensional  translation invariance by the presence
of the fivebrane. The sixteen spinors correspond to the breaking
of half of the thirty-two component supersymmetry of M-theory.  The
classical equations of motion  of the fivebrane in the absence of
fermions and background fields are~\cite{HSW}
\begin{equation}
G^{mn} \nabla_{m} \nabla_{n} X^{a'}= 0\ ,
\label{eqomone}
\end{equation}
and
\begin{equation}
G^{m n} \nabla_{m}H_{npq}  = 0.
\label{eqomtwo}
\end{equation}
where the worldvolume indices are $m,n,p=0,1,...,5$
and  the world tangent indices $a,b,c=0,1,...,5$.
The transverse indices are $a',b'=6,7,8,9,10$. The usual induced metric
and veilbien for a $p$-brane is given, in static gauge,  by
\bea
g_{mn} &=&
\eta_{m n}+\partial_{m}X^{a'} \partial_{n}X^{b'}\delta_{a' b'}\nonumber\\
&=& e^{\ a}_m \eta_{ab} e_{n}^{\ b}\ .\nonumber\\
\label{gdef}
\eea
The covariant derivative $\nabla$
is defined as the Levi-Civita connection with respect to the metric
$g_{m n} $.
The inverse metric $G^{mn}$ which also occurs 
is related to $g^{mn}$ by the equation
\begin{equation}
G^{mn} = {(e^{-1})}^{m}_{\ c} \eta ^{c a}
m_{a}^{\ d} m_{d} ^{\ b} {(e^{-1})}^{m}_{\  b}\ ,
\label{Gdef}
\end{equation}
where the matrix $m$ is given by
\begin{equation}
m_{a}^{\ b} = \delta_{a}^{\ b}
 -2h_{acd}h^{bcd}\ . 
\label{mdef}
\end{equation}
The field $H_{abc}$ is an anti-symmetric three-form and is the curl 
of $B_{ab}$. However is satisfies a non-linear self-duality constraint.  
To construct $H_{abc}$ we start from the three-form $h_{abc}$ 
which is self-dual;
\begin{equation}
h_{abc}=
{1\over3!}\varepsilon_{abcdef}h^{def}\ ,
\label{hsd}
\end{equation}
but it is not the curl of a three form gauge field. The field $H_{mnp}$
is then obtained as
\begin{equation}
H_{m n p}= e_{m}^{\ a}
e_{n}^{\ b} e_{p}^{\ c} {({m }^{-1})}_{c}^{\ d} h_{abd}\ .
\label{Hdef}
\end{equation}
Clearly, the self-duality condition on $h_{abd}$
transforms into a 
condition on $H_{m n p}$ and vice-versa 
for the Bianchi identify $dH=0$. 

\section{Soliton Dynamics}

In this section we wish to provide a complete derivation of the low energy
effective action for the wrapped M-fivebrane. We will review
the discussion given in~\cite{LWone} which treats the vector as well as
scalar modes and we refer the interested reader there for more details. 
It is possible to derive the Seiberg-Witten action relatively simply 
by considering the 
scalar modes alone and using $N=2$ supersymmetry to complete the 
action from only its 
scalar part~\cite{HLW}. 
We have chosen to discuss the complete analysis here because
the purely scalar argument is blind to many subtle and interesting 
features of the M-fivebrane, notably how the Abelian three-form can
reproduce the low energy effective action for non-Abelian vector fields.
In addition the argument presented below can be generalised to cases with
less supersymmetry. In these cases there is no a priori relation between the
vector and scalar dynamics. 
We also note that the  construction can  be performed 
in a manifestly
$N=2$ supersymmetric form~\cite{LWtwo} which perhaps best highlights 
the underlying geometry.

Our approach is to view the intersecting M-fivebrane configuration as
a threebrane soliton on a single  M-fivebrane worldvolume~\cite{three}.  Viewed
in this way the the soliton is a purely scalar field configuration of the
worldvolume theory and the Bogomol'nyi condition is just the 
Cauchy-Riemann equation for $s(z)$. From this point of view
we can obtain the low energy effective equations by
expanding the equations of motion to second order in derivatives
$\partial_{\mu}$, $\mu=0,1,2,3$ and field strengths $H_{mnp}$ around the 
threebrane background. 
To this order $h_{mnp}=H_{mnp}$ 
so that the field strength $H_{mnp}$ 
is  self-dual but with respect to the induced metric $g_{mn}$.
This implies that the ansatz in eq. \ref{Hdecomp} is 
incorrect as there are additional terms in $H_{mnp}$. In particular there
is a non-zero contribution to $H_{\mu\nu\lambda}$. Taking this into account
and expanding the equations of motion for two scalars (i.e. $X^6$ and $X^{10}$)
leads to the expressions
\be
E\equiv \eta^{\mu\nu}\partial_\mu\partial_\nu s
-\partial_z\left[{(\partial_\varrho s\partial^\varrho s)
\bar \partial\bar s\over(1+|\partial s|^2)}\right]
-{16\over(1+|\partial s|^2)^2}
H_{\mu\nu\bar z}H^{\mu\nu}{}_{\bar z}\partial \partial s
=0\ ,
\label{Escalar}
\ee
and
\be
E_{\nu}\equiv
\partial^\mu H_{\mu\nu z}-\partial_z\left[
{\bar \partial \bar
s\partial^\mu s\over(1+|\partial s|^2)}H_{\mu\nu  z}
-{\partial s\partial^\mu\bar s\over(1+|\partial s|^2)}H_{\mu\nu\bar z}
\right]=0\ .
\label{Evector}
\ee
In these expressions we have assumed that the threebrane soliton defined
by the Seiberg-Witten curve  $s(z)$ 
is dynamical due to its moduli $e_i$  becoming  $x^{\mu}$-dependent.   

Before reducing these equations to four-dimensions we need an ansatz for
the three-form components $H_{\mu\nu z}$. 
Let us restrict our attention here to the simplest case
of $N_c=2, N_f=0$, although the analysis can be generalised by considering
the appropriate curve. Explicitly,
eq. ~\ref{curve} leads to the curve 
\be
s = -{\rm ln}\left[ z^2 - u \pm \sqrt{(z^2-u)^2 - \Lambda}\right]\ ,
\ee
where $u = e_1e_2$ is the only moduli since $e_1+e_2=0$. 
Solving the self-duality 
condition requires  $H_{mnp}$  to take the form 
\be
H_{\mu\nu z} = \kappa {\cal F}_{\mu\nu}\ ,
\ee
where ${\cal F}_{\mu\nu} = F_{\mu\nu} + i*F_{\mu\nu}$ and  $\kappa$ 
is undetermined. The closure of $H_{mnp}$ requires that $\kappa$ 
is holomorphic.
We therefore write $\kappa = \kappa_0(x) \lambda_z$
where $\lambda_zdz = ds/du\ dz$ is the holomorphic one-form of the 
Riemann surface $\Sigma$. We are now free to choose $\kappa_0$ and we 
do this to
ensure that $F_{mn}$ is a closed two-form. We will see below that this 
in turn fixes
\be
\kappa = \left({da\over du}\right)^{-1}\lambda_z\ .
\ee
We have also introduced the periods 
\be
a = \int_A s dz\ ,\qquad a_D = \int_B s dz\ ,
\ee
where $A$ and $B$ are a basis of one-cycles of $\Sigma$, 
i.e. $s dz$ is identified with the Seiberg-Witten differential.
Note that the factor $(da/du)^{-1}$ normalises
the period of the form $\kappa dz$ to be one around the $A$-cycle. 
This reveals  another error in the argument
in section two as the choice of $\Lambda$ 
in eq. \ref{Hdecomp} will not lead to the Seiberg-Witten solution.

Finally to reduce these equations to four dimensions we project them over
a complete set of one-forms of $\Sigma$
\bea
0 &=& \int_{\Sigma} Edz\wedge\bar\lambda \nonumber\\
 &=& \partial^{\mu}\partial_{\mu}u I 
+ \partial_{\mu}u\partial^{\mu}u {dI\over du} 
- \partial_{\mu}u\partial^{\mu} u J
- 16{\bar {\cal F}}_{\mu\nu}{\bar {\cal F}}^{\mu\nu}
\left({d\bar a\over d\bar u}\right)^{-2}K \ ,\nonumber\\
0 & = & \int_\Sigma E_\nu dz\wedge \bar\lambda \nonumber\\
&=& \partial^{\nu}{\cal F}_{\mu\nu}\left({da\over du}\right)^{-1} I
-{\cal F}_{\mu\nu}\partial^{\nu}u{d^2 a\over du^2}
\left({da\over du}\right)^{-2}I 
+ {\cal F}_{\mu\nu}\partial^{\nu}u\left({da\over du}\right)^{-1}{dI\over du}
\nonumber\\
&&-{\cal F}_{\mu\nu}\partial^{\nu}u \left({da\over du}\right)^{-1}J
+{\bar{\cal F}}_{\mu\nu}\partial^{\nu}\bar u 
\left({d\bar a\over d\bar u}\right)^{-1}K
\ .\nonumber\\
\label{surfacetwo}
\eea
Here we encounter integrals over $\Sigma$ labelled by $I,J$ and $K$ 
and given below.
While it is straightforward to evaluate $I$ using the Riemann bilinear
relation the $J$ and $K$ integrals require a more sophisticated analysis.
This was done indirectly in~\cite{LWone} and directly in~\cite{LWtwo}
using properties of modular forms resulting in
\bea
I &\equiv&  \int_{\Sigma}\lambda\wedge\bar\lambda 
= {da_D\over du}{d\bar a\over d\bar u} - {da\over du}{d\bar a_D\over d\bar u} 
\ ,\nonumber\\
J &\equiv& R^2\Lambda\int_{\Sigma}\partial_z\left(
{\lambda_z^2\partial_{\bar z}\bar s\over 
1+R^2\Lambda\partial_z s\partial_{\bar z}\bar s}
\right)dz\wedge\bar\lambda\ = 0\ , \nonumber\\
K &\equiv&  R^2\Lambda \int_{\Sigma}\partial_{z}\left(
{\bar \lambda_{\bar z}^2\partial_{z}s\over 1
+R^2\Lambda\partial_z s\partial_{\bar z}
\bar s}\right)dz\wedge\bar\lambda =
-\left({d\bar a\over d\bar u}\right)^{3}{d{\bar \tau}\over d\bar a}
\ ,\nonumber\\
\label{IJK}
\eea
where $\tau = da_D/da$.
With these integrals we can now evaluate the four-dimensional equations of
motion
\bea
0&=&\partial^{\mu}\partial_{\mu}a(\tau-{\bar \tau}) + 
\partial^{\mu}a\partial_{\mu}a{d\tau\over da} 
+ 16 {\bar {\cal F}}_{\mu\nu}{\bar {\cal F}}^{\mu\nu}
{d {\bar \tau}\over d\bar a} \ ,\nonumber\\
0&=& \partial^{\nu}{\cal F}_{\mu\nu}(\tau-{\bar \tau})
+ {\cal F}_{\mu\nu}\partial^{\nu}u{d\tau\over du}
- {\bar{\cal F}}_{\mu\nu}\partial^{\nu}\bar u{d{\bar \tau}\over d\bar u} 
\ .\nonumber\\
\eea
Note that real part of the second equation is just 
the Bianchi identity for $F_{mn}$. 
Finally we see that these equations may be derived from the
Seiberg-Witten effective action
\be
S_{SW} = \int d^4 x\ {\rm Im} \left(  
\tau\partial_{\mu}a\partial^{\mu}{\bar a}
+16 \tau{\cal F}_{\mu\nu}{\cal F}^{\mu\nu}\right)\ .
\label{SWaction}
\ee

\section{Discussion}

In this review we have shown that by 
using brane dynamics one can obtain not only qualitative features of
quantum $N=2$ gauge theories such 
as the Seiberg-Witten curve
but also the precise details of the low energy effective action, including
instanton corrections. The example that we 
presented above also sets out a general method
which can be applied to configurations with less supersymmetry.
One first identifies a solitonic solution of the M-fivebrane and then the
low energy dynamics can be obtained from the M-fivebrane equations
of motion. The low energy effective action will contain scalar zero modes
from the moduli of the soliton and vector zero modes arising from 
the three-form $H_{mnp}$ and the brane topology.

Note that this construction is more subtle than the 
direct relation between
the geometry of the brane and the $\beta$-function that has been suggested. 
For example  the one-loop 
$\beta$-function coefficient  can be recovered from the 
bending of the branes. However
this interpretation has difficulties since 
if one identifies the coupling constant with the distance
between the two NS-fivebranes this gives a function 
$\tau(z)$ rather than $\tau(u)$. In addition
for $SU(N_c)$ gauge theories
there are in fact ${1\over 2}N_c(N_c-1)$ coupling constants 
and these cannot all be identified with the distance between the
two NS-fivebranes.\footnote{We are grateful to V. Khoze for this point.}

It is of course natural to see if M-theory can produce other details of
the Yang-Mills quantum field theory. For example it was pointed out in
reference~\cite{HLW} that the M-fivebrane also predicts an infinite number 
of higher
derivative terms which are
not holomorphic.
Although these terms are complicated one can easily see that they depend
upon the radius of the eleventh dimension and do not seem to agree with the
terms obtained from  quantum field theory\cite{O,LWone}. One can also consider
the M-fivebrane  prediction for low energy monopole dynamics
to see if this leads to the correct form for the monopole moduli space 
metric~\cite{mono}.

In closing let us mention some unsolved issues which warrant  further study. 
There
are other formulations of the M-fivebrane dynamics, which in addition
admit an action, and therefore it is natural to see if they can also 
reconstruct
the Seiberg-Witten effective action. However there is one immediate 
difficulty with using an action. Namely, since there is
no a priori distinction between $A$- and $B$-cycles, one expects that the
construction will be modular invariant. On the other hand, the Seiberg-Witten
action is not invariant under the $SL(2,{\bf Z})$ 
modular group, even though its equations of motion are.

Lastly there are
in fact discrepancies between the instanton coefficients predicted by
the Seiberg-Witten curves and those obtained by explicit calculations 
using instanton calculus for the cases $N_f=2N_c$~\cite{inst}. Perhaps
a more detailed consideration using  M-theory  will lead to alternative
forms for these curves.


\section*{References}
\begin{thebibliography}{99}
\bibitem{W}
E. Witten, \Journal{\NPB}{500}{3}{1997}, hep-th/9703166.
\bibitem{KLMVW}
A. Klemm, W. Lerche, P. Mayr, C. Vafa and N. Warner, 
\Journal{\NPB}{477}{746}{1996}, hep-th/9604034.
\bibitem{SW}
N. Seiberg and E. Witten, \Journal{\NPB}{426}{19}{1994}, 
hep-th/9407087;
\Journal{\NPB}{431}{484}{1994}, hep-th/9408099.
\bibitem{BBS}
K. Becker, M. Becker and A. Strominger, \Journal{\NPB}{456}{130}{1995}, 
hep-th/9507158.
\bibitem{V}
E. Verlinde, \Journal{\NPB}{455}{211}{1995}, hep-th/9506011.
\bibitem{HSW}
P.S. Howe, E. Sezgin and P.C. West, \Journal{\PLB}{399}{49}{1997}, 
hep-th/9702008.
\bibitem{HLW}
P.S. Howe, N.D. Lambert and P.C. West, \Journal{\PLB}{418}{85}{1998}, 
hep-th/9710034.
\bibitem{LWone}
N.D. Lambert and P.C. West, \Journal{\NPB}{524}{141}{1998}, 
hep-th/9710034.
\bibitem{LWtwo}
N.D. Lambert and P.C. West, \Journal{\PLB}{424}{281}{1998}, 
hep-th/9801104.
\bibitem{three}
P.S. Howe, N.D. Lambert and P.C. West, \Journal{\PLB}{419}{79}{1998}, 
hep-th/9710033.
\bibitem{O}
F. de Boer, K. Hori, H. Ooguri and Y. Oz, \Journal{\NPB}{518}{173}{1998}, 
hep-th/9711143.
\bibitem{mono}
N.D. Lambert and P.C. West, {\it Monopole Dynamics from the M-fivebrane}, 
hep-th/9811025.
\bibitem{inst}
N. Dorey, V.V. Khoze and M.P. Mattis, \Journal{\NPB}{492}{607}{1997}, 
hep-th/9611016.

\end{thebibliography}
\end{document}